\documentclass[lettersize,journal]{IEEEtran}
\usepackage{amsmath,amsfonts}
\usepackage{algpseudocode}
\usepackage{subcaption}
\usepackage{algorithm}
\usepackage{amssymb}
\usepackage{array}
\usepackage[caption=false,font=normalsize,labelfont=sf,textfont=sf]{subfig}
\usepackage{textcomp}
\usepackage{stfloats}
\usepackage{xcolor}
\usepackage{url}
\usepackage{verbatim}
\usepackage{graphicx}
\usepackage{cite}
\definecolor{mygreen}{rgb}{0.10,0.50,0.10}

\begin{document}

\title{Data-driven Under Frequency Load Shedding Using Reinforcement Learning}

\author{Glory~Justin,~\IEEEmembership{Student Member,~IEEE,}
 Santiago~Paternain,~\IEEEmembership{Member,~IEEE,}}



\maketitle

\begin{abstract}
Underfrequency load shedding (UFLS) is a critical control strategy in power systems aimed at maintaining system stability and preventing blackouts during severe frequency drops. Traditional UFLS schemes often rely on predefined rules and thresholds, which may not adapt effectively to the dynamic and complex nature of modern power grids. Reinforcement learning (RL) methods have been proposed to effectively handle the UFLS problem. However, training these RL agents is computationally burdensome due to solving multiple differential equations at each step of training. This computational burden also limits the effectiveness of the RL agents for use in real-time. To reduce the computational burden, a machine learning (ML) classifier is trained to capture the frequency response of the system to various disturbances. The RL agent is then trained using the classifier, thus avoiding multiple computations during each step of agent training. Key features of this approach include reduced training time, as well as faster real-time application compared to other RL agents, and its potential to improve system resilience by minimizing the amount of load shed while effectively stabilizing the frequency. Comparative studies with conventional UFLS schemes demonstrate that the RL-based strategy achieves superior performance while significantly reducing the time required. Simulation results on the IEEE 68-bus system validate the performance of the proposed RL method.
\end{abstract}


\section{Introduction}
\IEEEPARstart{P}{ower} system frequency is a reflection of the balance between power generation and load demand, thus serving as an important indicator of overall system stability \cite{b1}. However, power systems of today face added challenges that increase the complexity of maintaining steady frequency. Increasing interconnections of regional power grids for improved reliability of power supply has also led to dramatic increase in the size of power grids of today \cite{b2}. Modern power systems also experience a reduction in inertia due to a large share of renewable energy sources being interfaced with low inertia power electronics over high inertia conventional generators \cite{b3}. This reduction in overall system inertia has introduced new phenomena such as large frequency variations and severe system fluctuation after disturbances, which were previously unheard of in traditional power systems \cite{b1, b3}. These variations if not corrected, can lead to blackouts.

In cases of extreme frequency excursions, under-frequency load shedding (UFLS) schemes can be activated to minimize system separation or complete blackouts in an interconnected power system \cite{b4}. UFLS operates by reducing the amount of load connected to the system, thus ensuring power availability to critical loads and preventing escalation of system instability \cite{b5}. Conventionally, UFLS is achieved by shedding predefined blocks of loads once system frequency drops below a certain threshold \cite{b4, b5}. However, this method has a few drawbacks. Since the load shed does not take into account the disturbance rate, over or under-shedding is impossible to avoid. Loads can also be shed from safe areas instead of areas closer to the disturbance point. Unbalanced UFLS can also lead to over-loading of transmission lines. To prevent these occurrences, adaptive methods of UFLS have been considered to ensure that loads are shed in an optimal manner \cite{b4, b5}. These methods, however, are limited by the accuracy of the imbalance estimation.

With recent advancements in communucation and information technologies, more data is available for power system decision making and control from devices such as phasor measurement units (PMUs), advanced metering infrastructure (AMI) and wide area monitoring systems (WAMS) \cite{b6}. As a result, data driven methods such as machine learning are gaining prominence for application in power systems. One of these techniques is reinforcement learning (RL). RL is a branch of machine learning (ML) where agents are taught to take actions in an uncertain interactive environment and optimize performance based on feedback \cite{b6}. Due to its ability to adapt to complex new environments, learn model-free and fast decision making, it offers special advantages for real-time application to challenging problems \cite{b7}. This paper presents a data-driven approach to UFLS using RL. 

\subsection{Related Work}
Several methods for adaptive power system UFLS have been proposed in literature. Initial methods such as \cite{b8} used the initial slope of frequency deviation to determine the first step of UFLS, with subsequent steps determined later. In \cite{b9}, the authors go an extra step by first computing the frequency and rate of frequency change using time-domain simulations, then estimating the amount of load shed needed based on the disturbance type. While this method offers added advantages of giving the entire load amount beyond just the first step, it still requires large amounts of computation at each step, hence not practical for real-time application. In \cite{b10} and \cite{b11}, a distributed adaptive approach is proposed based on the rate of change of frequency. This approach, while offering the added distributed factor, also requires dynamic simulations to compute the system dynamics after each step of load shed. Many other dynamic simulation methods have also been proposed for UFLS. \cite{b12}  proposes an added constraint to ensure only the minimum amount of load is shed. \cite{b13} and \cite{b14} propose methods to correlate the amount of load shed needed with the size of disturbance and rate of change of frequency. The dynamic simulation methods while adaptive and promising, are limited in real-time applications due to the computational burdens they require.

To offer more computational efficiency for real-time applications, ML techniques have also been proposed for UFLS. In \cite{b1}, a combination of convolutional neural networks (CNN) and long short-term memory (LSTM) is used for to predict system frequency after a disturbance and provide optimal event-based UFLS. This method however, takes a centralized approach, not considering shedding on the areas most affected by the disturbances. In \cite{b2}, an integration of the system frequency response model (SFR) and an extreme learning machine (ELM) based model is used to predict frequency dynamics and UFLS. However, the addition of the model-based SFR increases the computational time requirements and effects of events on the load buses is not considered in the UFLS scheme. \cite{b15} uses a deep neural network combined with a risk-averse algorithm to predict load shed needed, however load is shed without considering the buses affected by the contingencies and their effects. For more adaptability and flexibility, RL methods have also been proposed for UFLS. 

RL is a powerful decision-making tool that has applied in multiple fields with tremendous success. Some of these fields include robotics \cite{b16}, autonomous driving \cite{b17} and playing games \cite{b18}. RL has shown the ability to capture hard-to-model dynamics, out-performing model-based methods in highly complicated tasks \cite{b6}. Leveraging this success in various fields, RL has also received more attention in recent years for solving power system problems.  RL has been used in research for prediction of uncertain conditions such as wind energy prediction in \cite{b19}. It has been proposed for control during normal operations such as automatic generation control in \cite{b20}. It is also used for power system security from false data attacks such as \cite{b21}. With these successes in a wide range of power system applications, RL is also proposed for UFLS. 

RL is a branch of ML where an agent is trained to make sequential decisions in an environment while maximizing cumulative reward. For problems with finite state/action spaces and lower complexity, the RL agent can learn the optimal policy by exploring all possible actions and choosing the actions of maximum value (value-based methods). However, for more complicated problems such as UFLS, the agent learns a parameterized policy that can select actions without consulting a value function (policy-based methods) \cite{b29}. One category of such methods is the actor-critic methods. In these methods, an actor network learns an approximation for the parameterized policy while a critic network learns an approximation for the state-value (Q) function which assesses the actions. This paper uses the Soft Actor-Critic algorithm, which is one of the actor-critic methods.

In \cite{b7}, the authors use a deep Q-network to provide real-time load-shedding decisions for improved voltage stability. The authors also use a convolutional long-short-term memory network to provide more information for a more descriptive reward function. In \cite{b23}, the authors use an RL algorithm for UFLS in a power system with high penetration of renewable energy. The agent is also shown to adapt to multiple events. In \cite{b24}, additional algorithms are included to improve the convergence speed of the RL algorithm and avoid violating system constraints by the RL agent. A knowledge-enhanced deep RL algorithm is used in \cite{b25} for intelligent event-based UFLs, where a linear decision space is designed and repeated actions are eliminated to reduce training time and enhance decision quality. In \cite{b26}, the authors use a deep deterministic policy gradient agent to effectively shed loads for stability correction.  

Safe RL \cite{b44} focuses on designing algorithms that ensure optimizing rewards while maintaining cost constraints. It aims to prevent the agent from taking actions that could lead to catastrophic outcomes. Constrained Markov Decision Processes (MDPs) \cite{b39} are a key framework in this domain, extending traditional Markov Decision Processes by incorporating constraints on the policies. For the UFLS problem, modeling as a  constrained MDP allows for the multiple power system constraints to be included in the optimization problem, thus ensuring safe decision making. Constrained MDPs have been applied in power systems for security constrained economic dispatch \cite{b42} and for emergency voltage control \cite{b43}.In this paper, they are used for the UFLS problem.

For the RL methods in \cite{b23}-\cite{b26}, with each action of the agent, multiple computations are required to determine the new state of the power system and the parameters for the reward function. The effects of partial observability are also not shown on the RL agents.  Targeted UFLS, shedding at unsafe areas over safe, to avoid unbalanced load shedding is also not considered in any of the above papers. In this paper, an RL approach is proposed for UFLS using data from PMUs. To reduce the computational burden and training time required for the agent, an ML classifier is introduced to analyze system frequency and determine new system states after each action. Both full and partial observability are considered. The agent is also trained to shed loads in a targeted manner, shedding at unsafe areas over safe areas to prevent unbalanced load shedding. The proposed method is shown to reduce training time and online application time by 1/100, over conventional methods.

\subsection{Main Contributions}
This paper presents an RL approach to event-based UFLS. The major contributions of this paper are as follows:
\begin{itemize}
    \item An RL agent is trained using the soft actor-critic (SAC) algorithm for UFLS at specific buses. Traditionally, this is done by shedding pocket of loads until the frequency is stabilized. However, this leads to over or under-shedding. Using a CMDP formulation, the RL agent is able to perform UFLS optimally while maintaining important safety constraints.
    \item To improve training efficiency and reduce computational burden, data-driven classifiers are introduced for assessing system frequency dynamics. 5 ML based classifiers are proposed and compared. These classifiers are trained to capture the system dynamics and detect unsafe or safe frequency conditions while training the RL agent. Using these classifiers, training time is reduced by over 90\%, thus improving training efficiency and preserving computational resources.
    \item To evaluate the robustness of the designed framework considering incomplete/partial observability of the PMUs, the effect of missing data on the ML classifiers is assessed. The RL agent trained using the GNN is shown to give a more accurate representation of the ground truth.
    \item Finally, the agent is shown to perform efficiently in a balanced manner, shedding load mainly in unsafe areas over safe areas to avoid unbalanced load shedding.
\end{itemize}

\section{Problem Description}
\begin{figure}
\centering
    \includegraphics[scale=0.4]{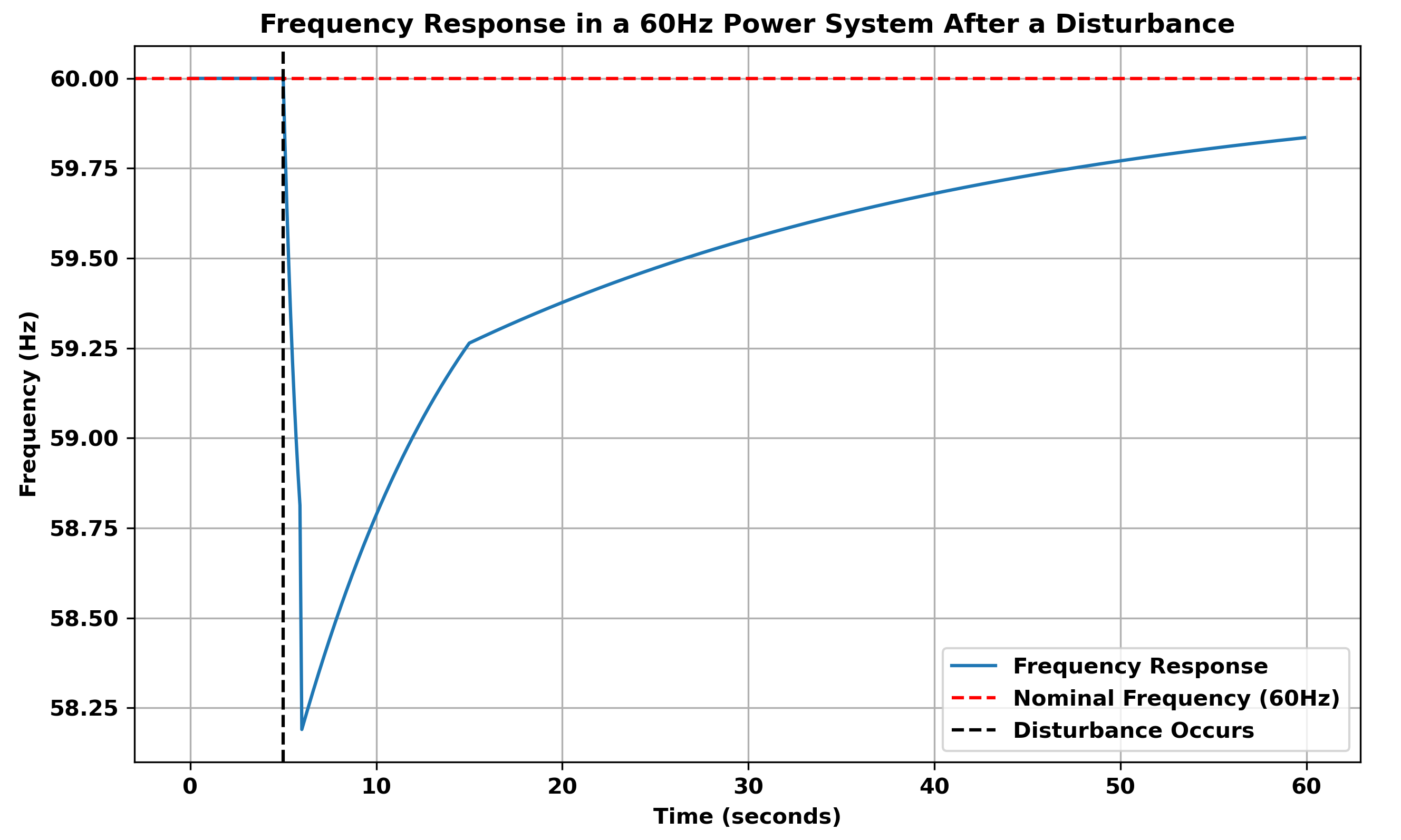}
    \caption{Frequency response after a disturbance at one of the generators in the IEEE 68-bus system. The disturbance here is a sudden increase in load $\Delta P$=2 MW at time $t=5$s. The frequency immediately drops following the disturbance and fails to recover to pre-contingency nominal frequency of 60Hz.}
    \label{fig:freq_response}
\end{figure}

For an interconnected power system with $N_G$ generators and $N_L$ loads in the system, the frequency dynamics at each generator can be represented by the swing equation. The frequency dynamics $f_i(t)$ of generator $i$ in the interconnected power system following a disturbance can be expressed by 
\begin{equation}\label{eqn:freq_dynamics}
    2H_i\frac{df_i(t)}{dt} = \Delta P_i - D_i\Delta f_i(t),\quad\forall i=1,\dots,N_G,
\end{equation}
where $\Delta f_i(t)$ represents the frequency deviation, $H_i$ is the generator inertia, $D_i$ is the load damping rate, $\Delta P_i$ is the active power imbalance caused by the disturbance for generator $i$ \cite{b27}. The imbalance power and grid dynamics are given by
\begin{equation}
    \Delta P_i = P_{mi}-P_{ei}
\end{equation}
\begin{equation}
    \begin{split}
        P_{ei} &= V_i\sum_{j=1}^N|Y_{ij}|V_j\cos{(\theta_i-\theta_j-\phi_{ij})},\\ &\forall i=1,\dots,N_G, j=1,\dots,N.
    \end{split}
\end{equation}
$\Delta P_i$ is obtained from the generator mechanical power $P_{mi}$ and the generator electrical power $P_{ei}$.  $P_{ei}$ is the electrical power obtained from the voltage results in the power flow, $V_i$ and $V_j$ are the bus voltage magnitudes at buses $i$ and $j$ respectively, $\theta_i$ and $\theta_j$ are the bus voltage angles at buses $i$ and $j$, $\phi_{ij}$ is the admittance angle for the line connecting buses $i$ and $j$, and $N=N_G+N_L$ is the total number of buses in the power grid network. The total power imbalance for the interconnected power system can be represented as 
\begin{equation}
    \Delta P_L = \sum_{i=1}^{N_G}\Delta P_i,\quad\forall i=1,\dots,N_G.
\end{equation}
A typical frequency response for a generator following a disturbance is shown in Fig. \ref{fig:freq_response}. The plot shown is for one of the generators in the IEEE 68-bus system. The generator has inertia constant $H=1$pu, damping rate $D=1$pu power/pu frequency following a sudden increase in load of size $\Delta P$=2 MW at time $t=5$s. The frequency immediately drops and fails to recover to pre-contingency nominal 60 Hz.

The most significant frequency index is the frequency nadir (FN). It represents the lowest point  ($f_{min}$ for positive $\Delta P_L$) or highest point ($f_{max}$ for negative $\Delta P_L$) in the frequency excursion post-contingency \cite{b27}. For the disturbance in Fig. \ref{fig:freq_response}, the FN is 2 Hz. To maintain stable operation and avoid damage to equipment, the value of FN must be kept within certain limits. Once these limits are exceeded, UFLS is triggered \cite{b27}.

The objective of UFLS is to return the frequency to acceptable limits while minimizing the amount of load shed to do so~\cite{b1}. When multiple contingencies are considered, the objective becomes to determine the optimal load-shedding solution for each contingency to satisfy the system frequency requirements. Formally let $N_g$ be the number of buses for load shedding, $P_{sj,max}$ the maximum load shed available at bus $j$, $\Delta P_L$ is the active power imbalance, $f_{min}$ and $f_{max}$ the minimum and maximum frequency threshold. We denote the system frequency dynamics in (\ref{eqn:freq_dynamics}) by  $\mathcal{F}(f)$. Then, the load-shedding problem can written as 
\begin{align}\label{eqn_opt_problem}
    \min_{P_{s}\in\mathbb{R}^{N_g}}  & \sum_{j=1}^{N_g} P_{sj}\\
    \text{s.t.}  & 0 \leq P_{sj} \leq P_{sj,max} \;\forall j=1,\ldots N_g\nonumber \\
    & f_{min} \leq f_i \leq f_{max} \;\forall i=1,\ldots N_G\nonumber \\
    & \mathcal{F}(f_i(t)) \;\forall i=1,\ldots N_G\nonumber.
\end{align}

According to \cite{b28}, for frequency below 60.5 Hz, generator limits are safe. However, for frequency below 59.5 Hz, UFLS is triggered to prevent generator trip. Therefore, in this paper, the minimum frequency is chosen as 59.5 Hz as done in \cite{b1} and \cite{b28} with a maximum limit of 60.5 HZ To avoid over-shedding.
 
\section{RL for UFLS}
UFLS is traditionally performed by shedding predetermined blocks of loads when the frequency drops below the threshold. Shedding predetermined loads, however, does not account for the size of the disturbance as well as other system dynamics. UFLS is a highly nonlinear problem with multiple constraints (see \eqref{eqn_opt_problem}). According to \cite{b25}, UFLs can be done in using off-line/on-line decision making and real-time implementation. In off-line decision making, transient stability simulation is done for the selected contingencies, then experts formulate and adjust the UFLS measures using trial and error from experience. However, due to too many operation scenarios and fault contingencies to consider in modern large power systems, off-line decision making has become time-consuming and computationally burdensome \cite{b25}. Optimization algorithms also face the challenge of too many constraints and controlled variables, thus facing convergence problems \cite{b25}. UFLS in a single step has also been considered. However, shedding large amounts in a single step can lead to over-shedding \cite{b10}. To avoid over-shedding, UFLS \eqref{eqn_opt_problem} is done in smaller steps \cite{b10}, thus making it a suitable problem for RL.

Changes in the power grid have necessitated the need for on-line decision making and controls \cite{b6}. RL has been proven to learn complex nonlinear relationships between input features and output actions directly from data \cite{b25}. Its abilities to adapt to a new environment and fast real-time decision speed offer many advantages for real-time decision making \cite{b7}.  Therefore, RL is used in this paper for on-line UFLS decision making.

\begin{figure}
\centering
    \includegraphics[scale=0.6]{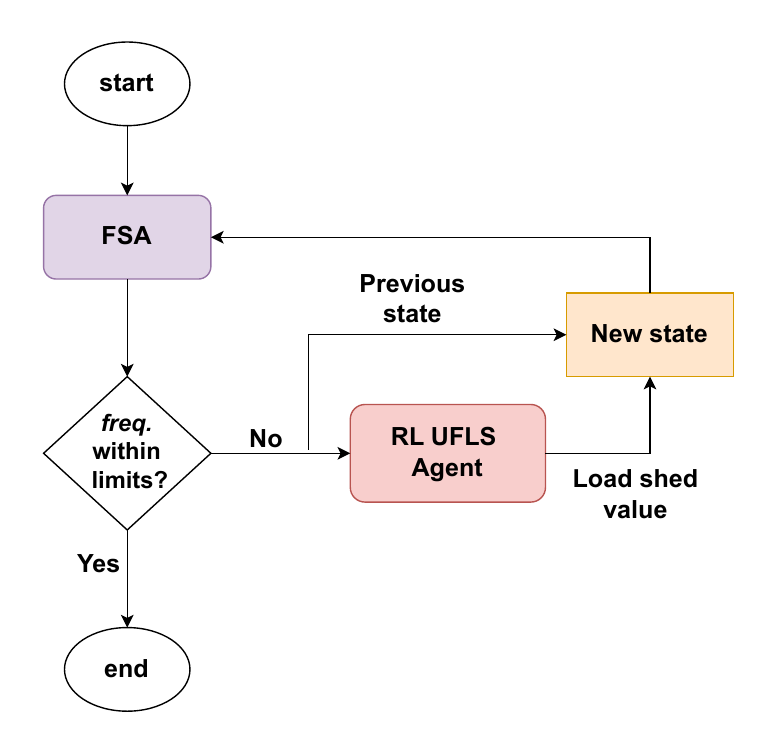}
    \caption{Illustration of the proposed UFLS scheme using RL. First, FSA is performed. If unsafe, UFLS is triggered. The agent outputs a set of load-shed coefficients. FSA is then repeated on the new state with the new load profile. If still unsafe, UFLS is repeated until safe.}
    \label{fig:RL_UFLS}
\end{figure}

An illustration of the proposed method is shown in Fig. \ref{fig:RL_UFLS}. First, FSA is performed. A set of contingencies are simulated and the operating condition is classified as safe if the minimum frequency $f_{min}>59.5$Hz and the maximum frequency $f_{max}<60.5$Hz. For operating conditions where the frequency requirements are not met for all contingencies, UFLS is required. The agent outputs a set of load-shed coefficients. FSA is then repeated on the new state with the new load profile. If still unsafe, UFLS is repeated until safe. To reduce the computational burden of FSA, an ML approach is proposed. 5 ML techniques are considered and compared. More details in Section \ref{sec:FSA_NN_description}.

\subsection{Modeling UFLS as a Constrained MDP.}
RL problems are modeled mathematically as a Markov Decision Process (MDP). An MDP is defined by a state space $\mathcal{S}$, an action space $\mathcal{A}$, a reward function that teaches the agent the objective $r(s,a):\mathcal{S}\times\mathcal{A}\rightarrow \mathbb{R}$, and the transition probability function $\mathbb{P}(.|s,a): \mathcal{S} \times \mathcal{A} \rightarrow \Delta (\mathcal{S})$ which maps a pair of state and action $(s,a)\in \mathcal{S} \times \mathcal{A}$ to a distribution in the state space \cite{b6}. At time step $k$, the agent at state $s_k$ selects an action $a_k$ based on a policy $\pi$. The environment gives the agent feedback based on the action so that the agent moves to state $s_{k+1}$, generated from the transition probability $\mathbb{P}(s_{k+1}|s_k,a_k)$ and receives reward $r(s_k,a_k)$. Let us define the expected infinite horizon discounted reward $J(\pi)$
\begin{equation}\label{eqn:J_pi}
    J(\pi) = \mathop{\mathbb{E}}_{\mathcal{T}\sim\pi}\left[\sum_{k=0}^{\infty} \gamma^k r(s_k,a_k,s_{k+1}) \right]
\end{equation}
where $\gamma \in [0,1)$ is the discount factor for the discounted long-term cumulative reward and $\mathcal{T} = (s_0,a_0,s_1,a_1,\dots)$ is a trajectory of states and actions \cite{b24}. In RL, the objective is to find the optimal policy $\pi^*$ that maximizes the cumulative cost. However, in the problem at hand we have constraints that need to be satisfied. As such, for each constraint associate a function $c_i(s_k,a_k,s_{k+1})$ with $i=1,\ldots, m$ where $m$ is the number of constraints. Analogous to \eqref{eqn:J_pi} define
\begin{equation}\label{eqn:C_pi}
    C_i(\pi) = \mathop{\mathbb{E}}_{\mathcal{T}\sim\pi}\left[\sum_{k=0}^{\infty} \gamma^k c_i(s_k,a_k,s_{k+1}) \right].
\end{equation}
Then the constrained MPD problem \cite{b39} consists of finding the policy that maximizes \eqref{eqn:J_pi} while keeping the constraints a above a certain level $\ell_i$
\begin{equation}
    \pi^* = \arg\max_{\pi} J(\pi) \;\mbox{s.t} \; C_i(\pi) \geq \ell_i \;\forall i =1,\ldots, m.
\end{equation}

In the context of UFLS, the states are the active $p_G \in \mathbb{R}^{N_G}$ and reactive power $q_G \in \mathbb{R}^{N_G}$ generation at each of the generation buses as well as the load active $p_L \in \mathbb{R}^{N_L}$ and reactive powers at each of the load buses. The training process for each time step is illustrated in Fig. \ref{fig:RL_UFLS}. The agent takes in states $s_k=[p_G,q_G,p_{L,k},q_{L,k}]$ at time step $k$ and outputs a set of buses for UFLS $a_k$ where $a_{k,i} = 0$ or $1\; \forall \;i=1,\dots,N_g$ at the selected load shed buses at each step. $a_{k,i}=1$ means to shed 5\% of load at bus $i$ at time step $k$ while $a_{k,i}=0$ means no load shed at bus i at time step $k$. After the action is executed, FSA is repeated for the same set of contingencies and the agent receives a reward $r_k$ based on the total load shed
\begin{equation}\label{eqn:reward_function}
    r_k = - \sum_{j=1}^{N_g} P_{sj,k},
\end{equation}
and it observes a constrained value based on the FSA results
\begin{equation}\label{eqn:constraint_function}
    c = \left\{ \begin{array}{ll}
        1 & \mbox{if $f_{min}>59.5$Hz and $f_{max}<60.5$Hz}\\
        0  & \mbox{if $f_{min}<59.5$Hz or $f_{max}>60.5$Hz}\end{array} \right.
\end{equation}
Thus, the agent has two objectives: minimizing power shed and satisfying the safety constraints $c$. Using a primal-dual parameter $\lambda$, we obtain a trade-off between both objectives, minimizing power shed and maximizing safety (see \eqref{eqn_dual_formulation}).

Notice that the level $\ell$ is related to the proportion of trajectories that are expected to be safe.Indeed, if $\ell=1$ the only possibility is that all trajectories satisfy the FSA requirements. This may be too ambitious and as such we set for a value strictly less than 1. The imbalance power $\Delta P_L$ is difficult to estimate due to the nonlinear relationship between the contingencies and the active power imbalance $\Delta P_i$ at each bus $i$. However, the frequency deviation is proportional to the active power imbalance $\Delta P_L$. For the constraints in \eqref{eqn_opt_problem}, the first constraint is satisfied by limiting the maximum number of steps the agent can take per episode. The second constraint is satisfied when the FSA requirements are met and the third constraint represents the frequency dynamics of the system which is used to perform FSA.

After the action is executed and FSA performed, if still unsafe, the agent takes in new states $s_{k+1}=[p_G,q_G,p_{L,k+1},q_{L,k+1}]$, where $p_{L,k}$ and $q_{L,k}$ are the new load active and reactive powers with 5\% load shed at the selected buses in $a_k$. This is illustrated in \eqref{eqn:new_state}
\begin{align}\label{eqn:new_state}
   p_{Lj,k+1} &= p_{Lj,k} - P_{sj} \\
   q_{Lj,k+1} &= q_{Lj,k} - q_{sj} \;\forall j=1,\ldots N_g.
\end{align}
$q_{sj}$ is the reactive power equivalent for the active power $P_{sj}$ shed at the load shed buses. The episode terminates after a maximum of $T$ steps or when the FSA requirement is fulfilled.

\subsection{Adapted SAC Algorithm}
We start this section by defining a single reward that can solve the CMDP problem. Indeed, CMDP have zero duality gap \cite{b40} and one can solve them in the dual domain as 
\begin{equation}\label{eqn_dual_formulation}
\min_{\lambda>0} \max_{\pi} 
\mathop{\mathbb{E}}_{\mathcal{T}\sim\pi}\left[\sum_{k=0}^{\infty} \gamma^k r_\lambda(s_k,a_k,s_{k+1}) \right],
\end{equation}
where $r_\lambda = r+\lambda c$. Intuitively $\lambda$ is a reward the agent receives for satisfying the FSA requirements $f_{min}>59.5$Hz and $f_{max}<60.5$Hz from the frequency dynamics in \eqref{eqn:freq_dynamics}. To find the appropriate value of $\lambda$ one can use primal-dual methods \cite{b41} or treat $\lambda$ as a hyperparameter. Since in this work effectively we have only one constraint, a hyperparameter search is tractable.

The SAC algorithm incorporates three key aspects: an actor-critic architecture with separate policy and value function networks, an off-policy formulation that allows use of previously collected data for efficiency, and entropy maximization to allow for exploration and maintain stability \cite{b30}. The entropy is used to encourage exploration and its importance can also be adjusted in the algorithm \cite{b24}. The objective function in (\ref{eqn:J_pi}) now becomes 
\begin{equation}
    J'(\pi) = \sum_{k=0}^K\mathbb{E}_{(s_k,a_k)\sim\mathcal{T}}\left[r(s_k,a_k) + \alpha \mathcal{H}(\pi(.|s_k)) \right],
\end{equation}
over $\mathcal{T}$, where $\mathcal{T}$ is the trajectory distribution induced by policy $\pi(a_k|s_k)$, $\mathcal{H}$ is the entropy of the policy $\pi$ at state $s_k$ and $\alpha$ is the temperature coefficient parameter which determines the significance of the entropy term \cite{b24, b30}. 

The state-action value function is given by
\begin{equation}\label{eqn:q_value_func}
   Q^{\pi}(s_k,a_k) = \mathbb{E}_{\pi} [r_k + \gamma Q^{\pi}(s_{k+1},a_{k+1})]. 
\end{equation}
On iteration, the state-action value converges to the optimum $Q^*(s,a)$. For the SAC algorithm, the soft Q-value function is used to assess the actions. This is similar to the Q-value function in (\ref{eqn:q_value_func}) but with an entropy term.  It is given by
\begin{equation}
    \mathcal{T}^{\pi}Q(s_k,a_k) \overset{\Delta}{=} r(s_k, a_k) + \gamma \mathbb{E}_{s_{k+1}\sim\mathbb{P}}\left[ V(s_{k+1}) \right],
\end{equation}
where the soft state value function $V$ is 
\begin{equation}\label{eqn:value_func}
    V(s_k) = \mathbb{E}_{a_k\sim\pi}\left[Q(s_k,a_k) - \alpha\log\pi (a_k,s_k)\right].
\end{equation}

As mentioned previously, large domains require approximations for policy iteration. Using function approximators, the networks can be trained using gradient descent instead of running evaluation and improvement to convergence \cite{b31}. Neural networks can be used to model the soft Q-function and the policy can be modeled as a Gaussian with mean and covariance obtained from neural networks \cite{b30, b31}. A parameterized soft Q-function $Q_{\theta}$ and policy $\pi_{\phi}(a_k,s_k)$ are considered, with parameters $\theta$ and $\phi$ respectively. The soft Q-function parameters are trained to minimize the soft Bellman residual

\begin{equation}
\begin{split}
    L_Q(\theta) &= \mathbb{E}_{(s_k,a_k)\sim\mathcal{T}}\\
    &\left[ \frac{1}{2} (Q_{\theta}(s_k,a_k)- (r(s_k,a_k) + \gamma\mathbb{E}_{s_{k+1}\sim\mathbb{P}}[V_{\Bar{\theta}}(s_{k+1})]))^2 \right].
\end{split}
\end{equation}
The value function $V$ is implicitly parameterized by the soft Q-function parameters in (\ref{eqn:value_func}). $L_Q$ can be optimized with stochastic gradients $\hat{\nabla}_{\theta}L_Q(\theta)$ where 
\begin{equation}
\begin{split}\label{eqn:Q_gradient}
  \hat{\nabla}_{\theta}L_Q(\theta) &= \nabla_{\theta} Q_{\theta}(a_k,s_k)(Q_{\theta}(s_k,a_k) - (r(s_k,a_k) \\
  &+ \gamma (Q_{\Bar{\theta}}(s_{k+1},a_{k+1} - \alpha\log(\pi_{\phi}(a_{k+1}|s_{k+1})))). 
  \end{split}
\end{equation}
The gradient update uses a target soft Q-function parameterized by $\Bar{\theta}$. $\Bar{\theta}$ is obtained as an exponential moving average of the soft Q-function weights $\theta$ \cite{b31}. 

In policy improvement, the policy is updated towards the exponential of the new soft Q-function \cite{b30,b31}.  This choice of update can be guaranteed to result in an improved policy in terms of its soft value. The most convenient projection is defined in terms of the Kullback-Leibler (KL) divergence \cite{b31}.  The policy $\pi$ is updated using the KL divergence
\begin{equation}
  \pi_{new} = \arg\min_{\pi'\in\Pi}\text{D}_{\text{KL}} \left(\pi'(.|s_k) \big| \big|\frac{\exp(\frac{1}{\alpha}Q^{\pi_{old}}(s_k,.)}{Z^{\pi_{old}}(s_k)} \right). 
\end{equation}
The partition function $Z^{\pi_{old}}(s_k)$ normalizes the distribution and does not contribute to the gradient with respect to the new policy, so can be ignored \cite{b30, b31}. The policy parameters can be learned directly by minimizing the expected KL-divergence 
\begin{equation}\label{eqn:sac_objective}
  J'_{\pi}(\phi) = \mathbb{E}_{s_k\sim\mathcal{T}}[\mathbb{E}_{a_k\sim\pi_{\phi}} [\alpha \log (\pi_{\phi} (a_k|s_k)) - Q_{\theta}(s_k,a_k)]].  
\end{equation}

Re-parameterizing the policy using a neural network transformation
\begin{equation}
    a_k = f_{\phi}(\varepsilon_k;s_k),
\end{equation}
where $\varepsilon_k$ is an input noise vector from some fixed distribution such as a spherical Gaussian. With this re-parameterization, the objective in (\ref{eqn:sac_objective}) can be rewritten as 
\begin{equation}
    J'_{\pi}(\phi) = \mathbb{E}_{s_k\sim\mathcal{T}, \varepsilon_k \sim \mathcal{N}}[\alpha\log\pi_{\phi} (f_{\phi}(\varepsilon_k;s_k)|s_k) - Q_{\theta}(s_k, f_{\phi}(\varepsilon_k;s_k))],
\end{equation}
where $\pi_{\phi}$ is defined implicitly in terms of $f_{\phi}$. The gradient $\hat{\nabla}_{\phi}J'_{\pi}(\phi)$ is approximated as 
\begin{equation}\label{eqn:pi_gradient}
\begin{split}
 \hat{\nabla}_{\phi}J'_{\pi}(\phi) &= \nabla_{\phi} \alpha \log(\pi_{\phi} (a_k|s_k)) + (\nabla_{a_k} \alpha\log(\pi_{\phi}(a_k|s_k))\\ 
 &- \nabla_{a_k} Q(s_k,a_k))\nabla_{\phi}f_{phi}(\varepsilon_k;s_k),
 \end{split}
\end{equation}
where $a_k$ is evaluated at $f_{\phi}(\varepsilon_k;s_k)$. The temperature parameter $\alpha$ is a hyperparameter that can also be optimized.

The SAC algorithm training process is shown in Algorithm \ref{alg:sac}. First, all the networks for parameterization are initialized: $\theta_1$ and $\theta_2$ for two critic networks, $\phi$ for the actor network and $\Bar{\theta}_1$ and $\Bar{\theta}_2$ for the target networks. Two critic networks are used to prevent positive bias in the policy environment \cite{b31}. For the UFLS problem, each iteration begins with an unsfe state $s_0$. The agent takes in an unsafe state $s_0\in\mathbb{R}^136$, where $s_0=[p_L,p_G,q_L,q_G]^T$ and outputs a set of $n=7$ load shed coefficients $a_0\in\mathbb{R}^n$. FSA is performed on the new state with load shed at certain buses $s_1$ and the agent receives a reward $r_(s_0,a_0)$ based on (\ref{eqn:reward_function}). If the new state is safe, the iteration ends and the actor and critic networks are updated using (\ref{eqn:pi_gradient}) and (\ref{eqn:Q_gradient}). The target network weights $\Bar{\theta}_1$ and $\Bar{\theta}_2$ are the same as the critic networks but updated fewer times to stabilize the training. If the new state $s_1$ is not safe, the UFLS is repeated with actions $a_k$ for a maximum of $K=4$ time steps. The iterations are repeated until convergence.

\begin{algorithm}
\caption{The SAC algorithm}\label{alg:sac}
\begin{algorithmic}
\State \textbf{Initialize} Q (critic) network parameters $\theta_1$, $\theta_2$, and policy (actor) network parameter $\phi$
\State \textbf{Initialize} target network weights $\Bar{\theta}_1 \leftarrow \theta_1, \Bar{\theta}_2 \leftarrow \theta_2$
\State \textbf{Initialize} empty replay buffer $\mathcal{T} \leftarrow \varnothing$
\For{each iteration}
    \For{each environment step}
        \State $a_k \sim \pi_{\phi}(a_k|s_k)$   \Comment{sample action from policy}
        \State $s_{k+1} \sim \mathbb{P}(s_{k+1}|s_k,a_k)$  \Comment{observe new state}
        \State $\mathcal{T}]\leftarrow \cup \left\{(s_k,a_k,r(s_k,a_k),s_{k+1}) \right\}$  \Comment{store transition}
    \EndFor
    \For{each gradient step}
        \State $\theta_i \leftarrow \theta_i - \lambda_Q  \hat{\nabla}_{\theta_i} L_Q(\theta_i)$ for $i\in \{1,2\}$ \Comment{update $\theta$}
        \State $\phi \leftarrow \phi - \lambda_{\pi}\hat{\nabla}_{\phi} J'_{\pi}(\phi)$  \Comment{update policy parameter}
        \State $\Bar{\theta}_i \leftarrow \tau \theta_i + (1-\tau) \Bar{\theta}_i$, for $i \in \{1,2\}$  \Comment{ update $\Bar{\theta}$}
    \EndFor
\EndFor
\State \textbf{Output: } $\theta_i, \theta_2, \phi$

\end{algorithmic}
\end{algorithm}

\subsection{FSA Using ML}\label{sec:FSA_NN_description}
\begin{figure}
\centering
    \includegraphics[scale=0.5]{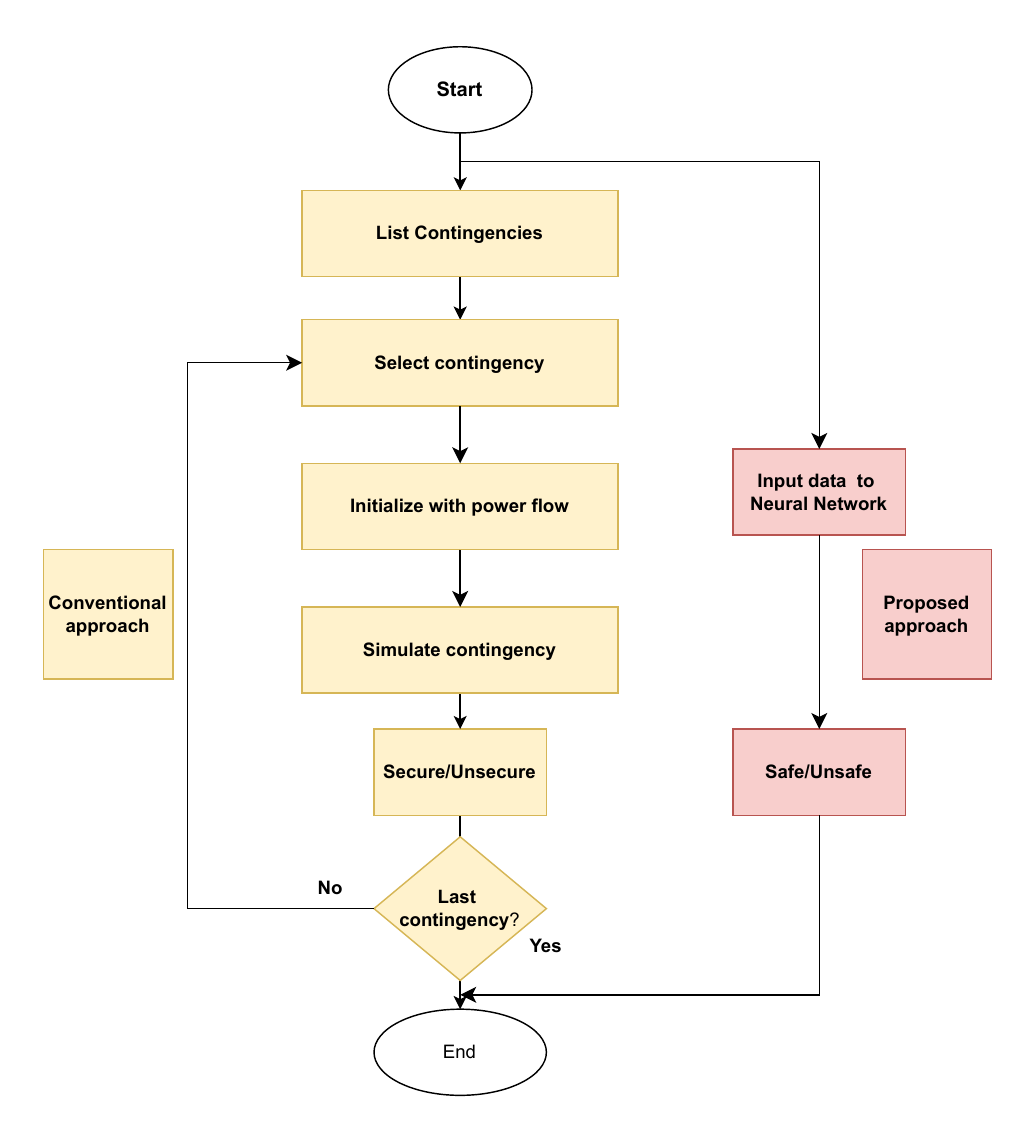}
    \caption{Neural network approach to FSA compared to the conventional method.}
    \label{fig:NN_FSA}
\end{figure}

To improve training efficiency and reduce the computational burden of training the RL agent, an ML approach is introduced for FSA. Similar approaches have been used in \cite{b37} for small signal stability and \cite{b38} for transient stability. Instead of simulating each contingency individually, a one-step process is used. The pre-contingency power flows are inputted to the ML classifier, which outputs 1 if safe for all contingencies or 0 otherwise. The operation of the ML approach over conventional time domain simulations is shown in Fig. \ref{fig:NN_FSA}, with the conventional approach in yellow and the ML approach in red. The data used to train the ML classifier is the pre-contingency bus voltages $v$, bus voltage angles $\delta$, net bus active powers $p$ and net reactive powers $q$. 5 ML techniques are compared for FSA: decision trees (DT), Support Vector Machines (SVM), Multilayer Perceptron (MLP), the Convolutional Neural Network (CNN) and the Graph Neural Network (GNN).

A GNN is a deep learning based method that operates in the graph domain. A graph is a data structure that models objects (nodes) and their relationships (edges). Graphs can be used to model the mathematical behavior of network topologies, with the data as signals on top of the graph. CNNs, as stated earlier, are only suitable for data in regular domains and struggle with irregular networked data. Graph neural networks (GNNs) use graph convolutions to incorporate the graph structure and topology information into the learning process. By incorporating this structure, important information about node dependencies and relationships is preserved, thus improving the training process \cite{b32}. 

The GNN used in this paper is the aggregation GNN \cite{b33}. Consider an undirected graph $\mathcal{G} = (\mathcal{V,E,W})$ with a set of $N$ nodes $\mathcal{V}$, a set of edges $\mathcal{E} \subseteq\mathcal{V}\times \mathcal{V}$ and an edge weight function $\mathcal{W}: \mathcal{E}\rightarrow\mathbb{R}_{+}$. For a power grid, the graph $\mathcal{G} = (\mathcal{V,E,W})$ represents a power grid with a set of N buses $\mathcal{V}$ and a set of transmission lines $\mathcal{E}\subseteq\mathcal{V}\times\mathcal{V}$. The graph is defined by an $N \times N$ real symmetric matrix $S$, termed the graph shift operator, where $S_{ij} = 0$ if ${(i,j)}\notin \mathcal{E}.$ To describe the operating conditions of the system, we assign a graph signal $x\in\mathbb{R}^N$, where $[x]_i$ represents the signal at node $i$. For the aggregation GNN, a single node is chosen and the graph signal is aggregated at this node. using successive communications between one-hop neighbours.  The aggregation sequence $z_i^f$ starts at the selected node $i$ where $f$ represents the states being considered, in this case voltage, active power and reactive power. The sequence is built by exchanging information with neighborhoods $S^kx^f$ and storing the result $[S^kx^f]_i$ at node $i$ for each $k$ from 0 to some number $N_{max}-1$. This results in the aggregation sequence 
\begin{equation}\label{eqn:agg_sequence}
  z_i^f = \left[[x^f]_i,[Sx^f]_i,\dots,[S^{N_{max}-1}x^f]_i\right]  
\end{equation} 
which can then be used to train a CNN \cite{b33}.

\section{Numerical Experiments}
Simulations are performed on the IEEE 68-bus system to verify the proposed method. The IEEE 68-bus system is a power system network with 68 buses, 86 transmission lines and 16 generators. The data is real data obtained from \cite{b34}. To generate multiple operating points, the generation at each bus is varied randomly from a uniform distribution between 70\% and 120\% of the base case data in \cite{b34}. The loads are also varied between 90\% and 150\%. This distribution is chosen to give a sufficient proportion of unsafe cases for testing the UFLS scheme. For each operating point, FSA is performed for a set of contingencies as described in Fig. \ref{fig:NN_FSA}. The frequency dynamics are analysed for each contingency and the operating point classified as safe if the post-contingency frequency is within limits for all contingencies. These contingencies are three-phase faults on select lines. This process generates 9950 data points with 65\% safe and 35\% unsafe. These data points are first used to train a CNN and GNN for FSA, then an RL agent for UFLS. This section is organized as follows: Section \ref{sec:FSA_results} shows the results of the ML methods for FSA, Section \ref{sec:full_observability} shows the results of the RL agent when FSA is done with full observability, Section \ref{sec:compare_efficiency} compares the performance of the GNN with other ML methods for FSA, Section \ref{sec:partial_observability} shows the effects of FSA in partial observability on the RL agent and Section \ref{sec:target_LS} shows the RL agent's performance in balanced load shedding, where load is shed in unsafe areas over safe areas. All simulations are done in Python 3.10 using an NVIDIA GeForce RTX 3070 GPU. The PyTorch library is used for all the trainings.

\subsection{FSA results}\label{sec:FSA_results}
5 ML classifiers are trained to perform FSA and reduce computational complexity. The input data for the DT, SVM, MLP and CNN is of size 68 X 4, containing the pre-contingency bus voltages $v$, bus voltage angles $\delta$, net bus active powers $p$ and net reactive powers $q$. The output is binary, 1 if safe for all contingencies and 0 otherwise. The CNN used has four consecutive layers of convolution and max pooling, connected to two fully connected layers, with the second fully connected layer as the output. All convolutional layers and the first fully connected layer use ReLU activation but the output layer has a Sigmoid activation for the output as a binary classification.  The GNN on the other hand is much smaller with only 3 layers. The input is an aggregation sequence generated using (\ref{eqn:agg_sequence}) for aggregation length $N_{max}=3$. Aggregating each of the four graph signals representing $v$, $\delta$ $p$ and $q$ for aggregation length $N_{max}=3$, gives an input of size 3 X 4.  Since the GNN input is much smaller, pooling is not required. The GNN has one convolutional layer, and two fully connected layers with the second as the output. The convolutional layer and first fully connected layer use ReLU activation while the output layer uses a Sigmoid activation, same as the CNN and MLP.  The architectures used for the CNN and GNN are similar to those used in \cite{b37} for small-signal stability assessment. All 5 ML classifiers are trained for 100 epochs with 0.001 learn rate and using the Adam optimizer \cite{b36}. DT and SVM are trained using Scikit-learn while the MLP, CNN and GNN are trained in PyTorch.

\begin{table}[htbp]
\caption{FSA results for the 5 different ML methods compared.}
\begin{center}
\begin{tabular}{|c||c|c|c|c|c|}
\hline
  & \textbf{Accuracy} & \textbf{Precision}&\textbf{Recall}&\textbf{Train time}&\textbf{Test time}\\
\hline
DT  & 93.70 & 95.14 & 95.24 & 47.2 & 0.9\\
SVM & 95.84 & 96.35 & 93.02 & 31.4 & 1.0\\
MLP & 94.84  & 94.73 & 95.06 & 150.1 & 1.1\\
CNN & 98.41  & 98.82 & 97.63 & 108.1 & 1.3\\
GNN & 94.24  & 91.85 & 98.94 & 103.0 & 0.9\\
\hline

\end{tabular}
\label{tab1:FSA_results}
\end{center}
\end{table}

For all 5 ML methods, the dataset is split into 75\% for training, 15\% for validation and 10\% for testing. The loss function $\ell$ is the binary cross entropy loss 
\begin{equation}\label{eqn:loss_func}
    \ell = -\frac{1}{M}\sum_{n=1}^M y_m\log(p(y_m))+(1-y_m)\log(1-p(y_m)),
\end{equation}
where $y_m=0$ or 1 is the label for data point $m$ and $p(y_m)\in\{0,1\}$ is the classifier output for data point $m$. $M$ is the total number of data points used for training. The results are summarized in Table \ref{tab1:FSA_results}. The training time and online testing time are in seconds. The CNN has the highest accuracy with 98.41\% while DT has the lowest accuracy with 93.70\%. The MLP however, takes the longest time to train while SVM trains the fastest. For real time testing, the CNN takes the most time with 1.3s while DT and the GNN take the least with 0.9s.

\subsection{UFLS in Full Observability}\label{sec:full_observability}
The next step is training the RL agent for UFLS. There are 16 generator buses and 33 load buses in the IEEE 68-bus system. Out of the 33 load buses, the 7 with the highest amount of load are selected for load shedding. The 3410 unsafe data points from the 9950 data points are used to train the agent. The agent is allowed a maximum of 4 steps. At each step, the agent selects some buses out of the 7 load-shed buses to shed 5\% of load. With this new load profile, FSA is repeated until the operating condition is safe for all the contingencies. This follows the process illustrated in Fig. \ref{fig:RL_UFLS}. 6 methods of FSA are used: Time Domain Simulations (TDS), DT, SVM, MLP, CNN and GNN.  The results shown are the average of 5 experiments with different random seeds for $\lambda=[0,1,2,10, 20]$. For $\lambda=0,$ the FSA constraints are not considered. The training is done for 10000 episodes using PyTorch. The maximum load shed amount at each bus is 20\%.

To train the agent, first the networks for the SAC algorithm are initialized. Four fully-connected networks are initialized: one actor network which represents the policy, two critic (Q) networks and one target network for stabilizing the training process. The target network has the same values as the critic 1 network but is updated less frequently. The actor network is updated using learn rate 1e-7 and the critic networks with learn rate 1e-6. All three networks are trained using the Adam optimizer. The networks are updated using the steps in Algorithm 1.  

\begin{figure}[h]
\centering
\includegraphics[scale=0.5]{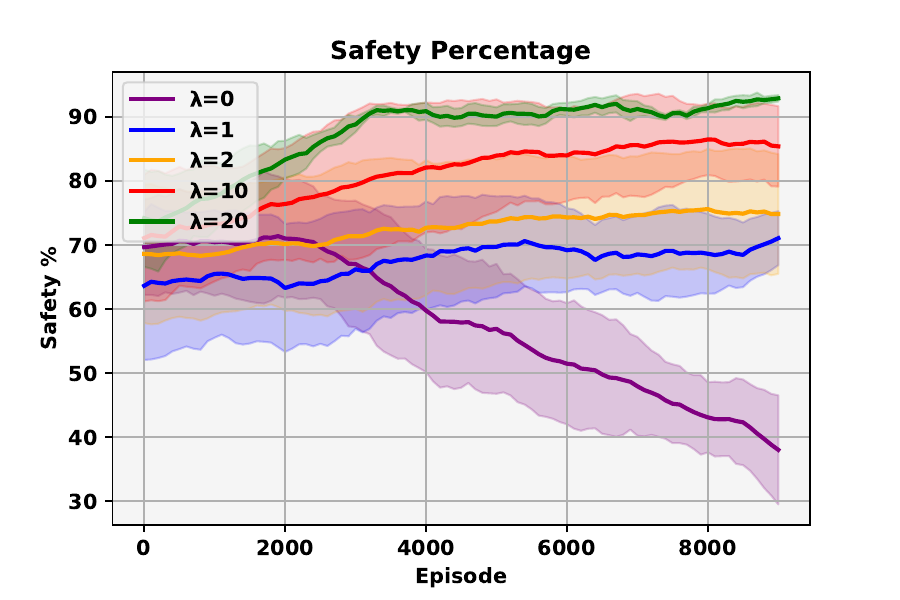}
\label{fig:unsafe_safe_GNN}
\caption{Safety percentage achieved by RL agent on 100 test data points during training. Safety progressively increases with increasing $\lambda$ with $\lambda=20$ giving the maximum safety percentage.}

\label{fig:unsafe_to_safe}
\end{figure}

During training, the RL agent is tested on 100 unsafe data points after 100 episodes, to evaluate the performance of the agent during training. The number of unsafe data points turned safe by the RL UFLS out of 100, gives the safety percentage. This safety percentage progress during training is plotted for the various values of $\lambda$ and shown in Fig. \ref{fig:unsafe_to_safe}. These results for the values of $\lambda$ are also summarized in Table \ref{tab1:GNN_Lambda_results}. The total power shed shows the total percentage of load shed at each of the 7 buses. From the results in Fig. \ref{fig:unsafe_to_safe}, $\lambda=20$ generates highest safety percentage while $\lambda=0$ generates the least number. For $\lambda=0$, the number reduces because the agent prioritizes minimizing load shed and FSA is not considered. Increasing $\lambda$ from 0 to 1 gives a drastic increase in safety of over 30\%, from a final average of 38.8 to a final average of 73.4\%. $\lambda=2$ takes the safety up to an average of 77.4\%, which is not as significant an increase as $\lambda=0,1$. Further increasing $\lambda$ from 2 to 10 gives a further increase in safety of over 10\%, from a final average of 77.4 to a final average of 88.5\%. $\lambda=20$ gives the maximum safety percentage of all the values of $\lambda$ considered, with an average of 92.2\%. 

\begin{table}[htbp]
\caption{Safety percentages achieved by the RL agent using the GNN for FSA and different values of $\lambda$.}
\begin{center}
\begin{tabular}{|c||c|c|c|}
\hline
  & \textbf{Test on classifier} & \textbf{Test on TDS}&\textbf{Total Load Shed}\\
\hline
$\lambda$=0  & 38.8 $\pm$ 17.3& 55.5 $\pm$ 15.5 & 0.32 $\pm$ 0.04\\
$\lambda$=1 & 73.4 $\pm$ 6.9 & 76.8 $\pm$ 4.5 & 0.48 $\pm$ 0.03\\
$\lambda$=2 & 77.4 $\pm$ 4.6 & 78.6 $\pm$ 6.1 & 0.52 $\pm$ 0.06\\
$\lambda$=10 & 88.5 $\pm$ 1.5 & 89.6 $\pm$ 1.4 & 0.58 $\pm$ 0.05\\
$\lambda$=20 & 92.2 $\pm$ 2.6 & 92.2 $\pm$ 0.8 & 0.63 $\pm$ 0.04\\
\hline

\end{tabular}
\label{tab1:GNN_Lambda_results}
\end{center}
\end{table}

\begin{figure}[h]
\centering
\includegraphics[scale=0.5]{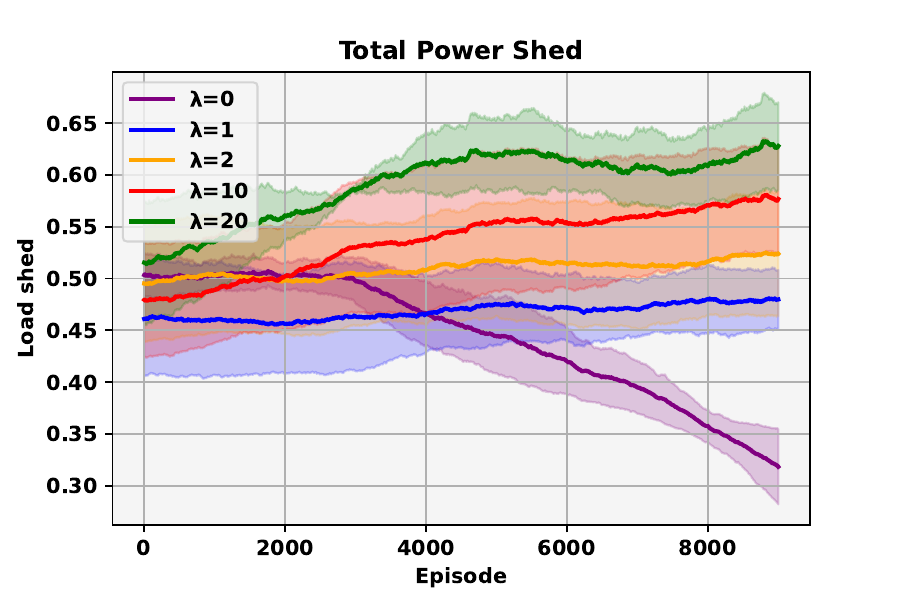}
\caption{Total power shed for 5 values of $\lambda$ using the GNN for FSA.  $\lambda=20$ shows the highest load shed for the GNN while $\lambda=0$ generates the least amount of load shed.}
\label{fig:total_load_shed}
\end{figure}

The total power shed for the for 5 values of $\lambda$ are also shown in Fig. \ref{fig:total_load_shed}. $\lambda=20$ gives highest load shed while $\lambda=0$ gives the least load shed. This is because with $\lambda=20$ the RL agent prioritizes safety over minimizing power shed, thus higher percentage of load is shed and higher safety percentage is obtained. With $\lambda=0$, the agent receives reward only for mininizing load shed, without considering FSA. Thus, $\lambda=0$ gives the least load shed but also the least safety percentage. 

\subsection{Computational Efficiency}\label{sec:compare_efficiency}
Next, the safety percentages as well as training and testing times are compared for 5 ML methods of FSA. The results are shown in Table \ref{tab1:RL_FSA_results}. Results shown are all for $\lambda=10$. Training and testing times are also compared with training and testing on Time Domain Simulations (TDS).

\begin{table*}[htbp]
\caption{RL agent performance compared for various ML methods. Safety percentages compared with 5 different FSA methods for $\lambda=10$. Training and testing times are compared.}
\begin{center}
\begin{tabular}{|c||c|c|c|c|c|}
\hline
  & \textbf{Test on classifier} & \textbf{Test on TDS} & \textbf{Total Load Shed} & \textbf{Training time (min)} & \textbf{Testing time (s)} \\
\hline
TDS  & 87.0 $\pm$ 2.4 & 87.0 $\pm$ 2.4 & 0.48 $\pm$ 0.05 & 7524 & 1171 \\
DT & 86.4 $\pm$ 1.5 & 83.8 $\pm$ 3.3 & 0.47 $\pm$ 0.07 & 44.1 & 7.4\\
SVM & 85.4 $\pm$ 1.6 & 83.6 $\pm$ 3.6 & 0.49 $\pm$ 0.07 & 46.6 & 6.5\\
MLP & 86.4 $\pm$ 0.8 & 91.2 $\pm$ 1.2 & 0.74 $\pm$ 0.17 & 72.9 & 8.3\\
CNN & 82.0 $\pm$ 2.2 & 76.0 $\pm$ 2.2 & 0.53 $\pm$ 0.11 & 48.1 & 7.7\\
GNN & 88.5 $\pm$ 1.5 & 89.6 $\pm$ 1.4 & 0.63 $\pm$ 0.04 & 48.6 & 5.5\\
\hline

\end{tabular}
\label{tab1:RL_FSA_results}
\end{center}
\end{table*}

From the results in Table \ref{tab1:RL_FSA_results}, the ground truth (TDS) gives safety percentage of 87 out of 100 unsafe test data points. It however, takes the highest amount time to train with 7524 minutes (125 hours) and requires the highest amount of time for real-time testing with 1171 seconds (20 minutes). DT reduces this training time by 99.9\%, training in an average of 44 minutes. DT achieves a safety percentage of 86.4\% which gives 83.8\% on testing with the ground truth. DT is also more efficient for real-time application than TDS, requiring only 0.6\% of the time required by TDS for real-time testing. Using SVM requires slightly more time than DT for training the RL agent, with 46.6 minutes. It however, performs faster in real-time testing requiring 6.5 minutes. The RL agent with SVM FSA is also able to achieve 85.4\% safety which translates to 83.6\% on testing with the ground truth for FSA. Training with an MLP for FSA gives the longest training time of the ML techniques, with 72.9 minutes. It also requires the most time for real time testing. The CNN achieves the lowest safety percentage with 82.0\%. Testing on TDS shows 76.0\%, a difference of 6\%. This shows lower accuracy and thus lower generalization potential for the CNN. The GNN, however, shows the highest accuracy with a difference of 1.1\% between the test safety results and TDS test safety results. The GNN thus shows the best generalization ability of the 5 ML methods compared. The GNN also shows the best performance for real-time testing requiring 5.5s. It is thus also the most efficient for real-time testing. All 5 ML techniques however, give significant computational benefits over TDS. The results shown are for FSA in full observability using data from PMUs.

\subsection{UFLS in Partial Observability}\label{sec:partial_observability}
The results shown in Section \ref{sec:full_observability} are for a data-driven approach to FSA, where the ML classifiers are trained on PMU data to identify safe and unsafe operating conditions. It is assumed that data is available at all the buses in the system. However, that is rarely the case. Missing PMU data is unavoidable. This section analyses the effect of missing data at certain buses in the system on the ML techniques for FSA. The trained RL agents from Section \ref{sec:full_observability} are evaluated for FSA with missing data at 25\% of the buses in the IEEE 68-bus system. The effects of this missing data and reduced accuracy are shown in Table \ref{tab1:RL_FSA_missing}. From the results, we see that the RL agents trained on the various ML FSA methods show a difference in predictions of 4.3\% with DT up to 28.3\% with the CNN. The GNN on the other hand, changes by 0.3\% in safety percentages with missing data. This shows that the GNN is unaffected, producing comparable results with the tests on ground truth with or without missing data.

\begin{table*}[htbp]
\caption{RL agent results with missing data at 25\% of the buses.}
\begin{center}
\begin{tabular}{|c||c|c|c|}
\hline
  & \textbf{Test with missing data}  & \textbf{Test without missing data} & \textbf{Test on TDS}\\
\hline
DT & 90.7 $\pm$ 1.8 & 86.4 $\pm$ 1.5 & 83.8 $\pm$ 3.3 \\
SVM & 99.0 $\pm$ 0.0 & 85.4 $\pm$ 1.6 & 83.6 $\pm$ 3.6  \\
MLP & 92.4 $\pm$ 1.4 & 86.4 $\pm$ 0.8 & 91.2 $\pm$ 1.2 \\
CNN & 53.7 $\pm$ 5.0 & 82.0 $\pm$ 2.6 & 76.0 $\pm$ 2.2\\
GNN & 88.2 $\pm$ 0.4 & 88.5 $\pm$ 1.5 & 89.6 $\pm$ 1.4\\
\hline

\end{tabular}
\label{tab1:RL_FSA_missing}
\end{center}
\end{table*}

\subsection{Balanced Load-Shedding}\label{sec:target_LS}
To prevent unbalanced load shedding, it is important for the RL agent to shed more loads at unsafe areas over safe areas. Unguided, the RL agent sheds maximum load at the unsafe contingency bus for 7360 out of 1000 test cases. This shows that the RL agent instinctively sheds loads in a balanced manner, prioritizing unsafe areas over safe, thus avoiding tie-line overloads and unbalanced shedding.

\section{Conclusion}
This paper presents an RL approach to UFLS using a CMDP formulation. By including constraints in the MDP formulation, the agent is able to achieve the UFLS objective without violating crucial system constraints. With $\lambda=20$ we are able to achieve 92.2\% safety while ensuring minimal loadshed. To reduce the computational burden, an ML classifier is used for FSA, analyzing system frequency after each training step for the RL UFLS agent. Using the ML classifier, training time is reduced by up to 0.6\% of the time required using time domain simulations. Comparing 5 ML classifiers for FSA, we obtain fastest real-time performance with the GNN as well as robustness to missing data. The RL agent is also shown to shed load in a balanced manner, prioritizing shedding in unsafe contingency areas over safe areas.


\begin{IEEEbiographynophoto}{Glory Justin}
received the Bachelor of Engineering degree in electrical and electronic engineering from Federal University of Technology Owerri, Imo State, Nigeria, in 2017 and the M.S. degree in electrical engineering from Rensselaer Polytechnic Institute, Troy New York, in 2022. She has done summer internships with Heliogen as an Engineering Intern in 2022 and GE Aerospace research as an Electric Machines and High Voltage Engineering Intern in 2023. She is currently pursuing the Ph.D. degree in electrical engineering at Rensselaer Polytechnic Institute, Troy, NY. Her research interests include power system stability and control, and machine learning
\end{IEEEbiographynophoto}

\begin{IEEEbiographynophoto}{Santiago Paternain}
received the B.Sc. degree in electrical engineering from Universidad de la República Oriental del Uruguay, Montevideo, Uruguay in 2012, the M.Sc. in Statistics from the Wharton School in 2018 and the Ph.D. in Electrical and Systems Engineering from the Department of Electrical and Systems Engineering, the University of Pennsylvania in 2018. He is currently an Assistant Professor in the Department of Electrical Computer and Systems Engineering at the Rensselaer Polytechnic Institute. Prior to joining Rensselaer, he was a postdoctoral Researcher at the University of Pennsylvania. He was the recipient of the 2017 CDC Best Student Paper Award and the 2019 Joseph and Rosaline Wolfe Best Doctoral Dissertation Award from the Electrical and Systems Engineering Department at the University of Pennsylvania. His research interests lie at the intersection of machine learning and control of dynamical systems.
\end{IEEEbiographynophoto}

\vfill

\end{document}